\documentclass[doublecol]{epl2}

\title{Transversal interface dynamics of a front connecting a
stripe pattern to a uniform state}
\shorttitle{Transversal interface dynamics of a front}

\author{Marcel G. Clerc \inst{1} \and Daniel Escaff \inst{2}
\and Ren\'{e} Rojas\footnote{E-mail address: rrojas@dfi.uchile.cl} \inst{1}}

\institute{ \inst{1} Departamento de F\'{i}sica, FCFM,
Universidad de Chile, Casilla 487-3,
Santiago, Chile,\\
\inst{2} Complex Systems Group, Facultad de Ingenier\'\i a,
Universidad de los Andes, Av. San Carlos de Apoquindo 2200,
Santiago, Chile. }


\abstract{Interfaces in two-dimensional systems exhibit unexpected
complex dynamical behaviors, the dynamics of a border connecting a
stripe pattern and a uniform state is studied. Numerical simulations
of a prototype isotropic model, the subcritical Swift-Hohenberg
equation, show that this interface has transversal spatial periodic
structures, zigzag dynamics and complex coarsening process. Close to
a spatial bifurcation, an amended amplitude equation and a
one-dimensional interface model allow us to characterize the
dynamics exhibited by this interface.}

\begin{document}

\maketitle

\section{Introduction}Non equilibrium processes often lead in
nature to pattern formation developing from a homogeneous state
through the spontaneous breaking of the symmetries present in the
system \cite{Prigogine}. In recent decades, much effort has been
devoted to the study of pattern formation (see review \cite{Cross}
and the references therein) arising in systems such as chemical
reactions, gas discharge systems, CO$_{2}$ lasers, liquid crystals,
hydrodynamic or electroconvective instabilities, and granular matter
(see review \cite{Aranson}), to mention a few. A unified description
for the dynamics of spatially periodic structures, developed at the
onset of a bifurcation, is achieved by means of amplitude equations
for the critical modes. Such a description is valid in the case of
weak nonlinearities and for a slow spatial and temporal modulation
of the base pattern \cite{Cross}. As an example, the
Newell-Whitehead-Segel equation \cite{NewellWhitehead} describes the
dynamics of a stripe pattern formed in a two-dimensional system.
Another ubiquitous phenomenon in nature is the interface dynamics or
{\it front propagation}. The concept of front propagation, emerged
in the field of population dynamics
\cite{Fisher,Kolmogorov,ClercEscaffKenkre}, has gained growing
interest in biology, chemistry, physics, and mathematics (See,
e.g.\cite{Murray} and references therein). These interfaces connect
two extended states, such as: uniforms states, patterns,
oscillatory, standing waves, spatio-temporal chaotic and so forth.

In one-dimensional systems, an interface connecting two uniform
stable states, the most favorable state---for instance
energetically---invades the other one with a constant and unique
speed \cite{ClercResidori,ResidoriPetrossian}. This speed is zero,
that is the front is motionless, at the Maxwell point \cite{Pomeau}.
The above picture changes, when one considers an interface
connecting a pattern state and a uniform one or two patterns. Due to
spatial translation symmetry breaking, the interface is motionless
in a range of parameters, \emph{the pinning range}
\cite{Pomeau,Clerc2005,Aranson2000}. This behavior is called locking
phenomenon or pinning effect. In bidimensional dynamical systems,
few experimental and theoretical studies have been performed on
fronts connecting patterns and uniform states
\cite{MeronElphick2006,MalomedNepo1991,AhlersFront,Taki}.

The aim of this letter is to study the dynamical behaviors of a
front connecting a stripe pattern to a uniform state. Numerical
simulations of a prototype model---the isotropic Swift-Hohenberg
equation---show that the locking phenomenon of a flat interface
persists. However, this flat interface is nonlinearly transversely
unstable, that is, a finite perturbation of this interface leads to
the appearance of an initial wave number which is subsequently
replaced by zigzag dynamics, which presents a complex coarsening.
Increasing the longitudinal interface size, the flat interface
exhibits a transversal spatial instability, which originates a
periodical structure at the interface. We have termed these
interfaces \textit{embroideries}. In order to explain these
behaviors in an unified manner, we make use of the amended
Newell-Whitehead-Segel equation and prototype one-dimensional model
for the interface, which describe properly the dynamics observed on
the interface connecting a stripe pattern with a uniform state.

\section{The model}A simple isotropic model which exhibits coexistence between a
uniform state and a stripe pattern is (the subcritical
Swift-Hohenberg equation \cite{Cross})
\begin{equation}
\partial _{t}u=\varepsilon u+\nu u^{3}-u^{5}-\left( \vec{\nabla}
^{2}+q^{2}\right) ^{2}u,  \label{E-SubcriticaSH}
\end{equation}
where $u\left( x,t\right)$ is an order parameter, $\varepsilon$ is
the bifurcation parameter, $q$ is the wave number of the stripe
pattern, $\nu $ the control parameter of the type of bifurcation
(supercritical or subcritical), and $\vec{\nabla}^{2}$ is the
Laplacian operator. The model (\ref{E-SubcriticaSH}) describes the
confluence of a stationary and a spatial subcritical bifurcations
with reflection symmetry ($u\rightarrow -u$), when the parameters
scale as $u\sim \varepsilon ^{1/4}$, $\nu \sim \varepsilon ^{1/2}$,
$q\sim \varepsilon ^{1/4}$, $\partial _{t}\sim \varepsilon $ and
$\vec{\nabla}\sim \varepsilon ^{1/4}$ ($\varepsilon \ll 1$). The
above model is often employed in the description of patterns
observed in Rayleigh-Benard convection \cite{Cross}. For small and
negative $\varepsilon $ and $-9\nu ^{2}/40\equiv \varepsilon
_{sn}<\varepsilon <0$, the system exhibits coexistence between a
uniform state $u(x,y,t)=0$ and a stripe pattern $u\left(
x,y,t\right) =\sqrt{\nu } \left( \sqrt{2(1+\sqrt{1+40\varepsilon
/9\nu })}\cos \left( \vec{q}\cdot \vec{r}\right) \right) +o(\nu
^{5/2})$, where $\vec{q}$ is an arbitrary vector with modulus $q$.
For $\varepsilon =\varepsilon _{sn}$ the model has a saddle-node
bifurcation that give rises to stable and unstable patterns. For
$\varepsilon =0$ the uniform state becomes unstable. When one
considers the above model in one spatial dimension, it is well-known
that in the coexistence region ($\varepsilon _{sn}<\varepsilon <0$)
the model exhibits a front connecting a spatially periodic solution
and uniform state. For $ \varepsilon $ close to $0$ ($\varepsilon
_{sn}$), the pattern (uniform) state invades the uniform (pattern)
state, i.e. the system displays front propagation. The front is
motionless at the interval $\varepsilon _{-}<\varepsilon
<\varepsilon _{+}$, the pining range \cite{Pomeau}. Note that, in
this range, the state which is energetically more favorable does not
invade the less favorable one. It is important to note that the
interface has two characteristic lengths, the pattern length
$2\pi/q$ and the interface size, which is represented by $\lambda$
in the inset of Fig.~\ref{F-fig1}. Decreasing $\nu$ the interface
size increases.

\section{Numerical results} In two spatial dimension the above scenario
changes drastically. If one considers a similar parameters setup
($\varepsilon _{sn}<\varepsilon _{-}<\varepsilon <\varepsilon
_{+}<0$), the flat interface of model (\ref{E-SubcriticaSH})
exhibits locking phenomenon (cf. Fig.~\ref{F-fig1}a). However, when
the longitudinal interface size $\lambda$ is increased the interface
suffers a supercritical transversal spatial instability and it gives
rise to transversal periodic structures. The typical observed
structures are depicted in Fig.~\ref{F-fig1}. We term these
periodical structures at the interface \textit{embroidery}. The
embroidery size is proportional to the longitudinal interface size.
The embroideries are consequence of two facts: the spatial isotropy
and the interaction of enveloped variation with the underlying
pattern (pinning effect).  Seeing that at the interface the stripe
pattern can develop in any direction as consequence of the isotropy,
however the interaction of enveloped variation with the underlying
pattern freezes the growth of these unstable modes, as we shall see
with the amended amplitude equation. Based on the 1D theory of front
interaction which predicts a family of localized patterns
\cite{ClercFalcon2005}, one expects to find a family of localized
stripes, that is, the localized stripes are the transversal
expansion of 1D localized patterns. The typical stable localized
states observed in model (\ref{E-SubcriticaSH}) are illustrated in
Fig. 2. It is important to note that, localized stripe patterns with
flat interfaces have been observed in ref. \cite{BrandSakaguchi} .

Numerically, we observe that in a finite region of
parameters the embroidery interface is linearly stable, i.e. this
interface with a small initial perturbation evolves to the
interface without perturbations. However, when we consider a
large perturbation  the interface does not evolve to the
interface without perturbation. This interface exhibits  a
nonlinear zig-zag instability, which is characterized by a
transversal instability without a well-defined wavelength. In
spite of this, initially this instability has a well defined wave
number close to $2\pi/q$, later on, the sinusoidal interface
becomes an angled line composed of pieces of interface turned with
well define angles, zig-{\it facet or zag-facet}. It is
important to note that the flat interface--in the regime of
parameters where it is linearly stable---exhibits the same
behavior, that is, a finite initial perturbation of the flat
interface gives rise to a zig-zag dynamics. In order to illustrate
this nonlinear instability, Fig.~\ref{F-fig3}a shows two
embroidery interfaces, one with a initial small perturbation
(left interface) and the other with a finite one. After a finite
time, the interface with initial small perturbation evolves to an
embroidery interface. While the Right-interface develops a zig-zag
structure and the pattern propagates over the uniform state (cf.
Fig.~\ref{F-fig3}b). Hence, the flat and embroidery interfaces are
nonlinearly transversely unstable.

In the zig-zag interface, two adjacent facets whose orientations
are opposite, are connected by a region of strong curvature that
we term {\it a corner}. The dynamics shown by the
zig-zag interface consists then in reassembling domains
of even orientation, the angle facets staying unchanged, which is
a {\em coarsening dynamics}. This process occurs due to
annihilations of corners and without a characteristic length
scale. Actually, the averaged domain size increases regularly in
time. Simultaneously to this coarsening process, the zigzag
interface propagates from one extended state to the other one.
Figure 4 depicts the typical coarsening process
observed at the interfaces.

\section{The amended amplitude equation}To explain the dynamics
exhibited by the interfaces of model (\ref{E-SubcriticaSH}) in a
unified framework, close to  the spatial bifurcation ($
\varepsilon \ll 1$, and $\nu \sim \varepsilon ^{1/2}$), we can
introduce the ansatz
\begin{eqnarray}
u(x,y,t)=a_{0}A\left( X=\frac{x}{l_{0}}, Y=\frac{y}{\sqrt{q
l_{0}}},\tau=\frac{t}
{t_{0}}\right)  e^{iqx}\nonumber\\
-5\frac{a_{0}^{5}|A|^{2}A^{3}}{8q^{2}}e^{i3qx}-
\frac{a_{0}^{5}A^{5}}{24q^{2}
}e^{i5qx}+\cdots+c.c.,\label{E-Ansatz}
\end{eqnarray}
where $a_{0}\equiv$ $\sqrt{3\nu/10}\varepsilon^{1/4}$,
$l_{0}=2q\sqrt {10}/3\sqrt{\left|  \varepsilon\right|  }$,
$t_{0}=10/9\nu^{2}\left| \varepsilon\right|  $, and $A(X,Y,\tau)$ is
the envelope of the pattern that describes the front solution (when
the envelope is uniform and not null the initial model~(\ref
{E-SubcriticaSH}) has an stripe pattern in the $x$-direction),
 the "$\cdots$" stands for the high order terms in the
envelope $A$, $c.c.$ means complex conjugate, and $\left\{
X,Y,\tau \right\} $ are slow variables. In this ansatz
(\ref{E-Ansatz}), we consider that $q$ is of order one or larger
than the other parameters. Introducing the above ansatz in
equation (\ref{E-SubcriticaSH}), we find the following solvability
condition for the envelope (\textit{Amended subcritical
Newell-Whitehead-Segel equation})

\begin{equation}
\partial_{\tau}A=\left\{  \epsilon+\left|  A\right|  ^{2}-\left|  A\right|
^{4}+\left(  \partial_{X}-\frac{i\partial_{YY}}{2q}\right)
^{2}\right\} A+\eta A^{3}e^{i\kappa x},
\label{E-NewellWhiteheadSegel}
\end{equation}
where $\epsilon \equiv -10\varepsilon /9\nu ^{2}$,
$\kappa=2\sqrt{10}q/3\nu\sqrt{\varepsilon}$, and $\eta \equiv 1/3 $
for model (\ref{E-SubcriticaSH}). Using the above scaling the terms
inside brackets are order $\epsilon$ \cite{Budd}, and the last term
is exponentially small which is regularly neglected in the multiple
scaling approach. However, to account for the coupling between the
large scale envelope and the small scale underlaying the stripe
state, we consider $\eta$ as free parameter and $\kappa$ a finite
number. Notice that consider other solvability conditions for the
envelope one can obtain diverse small exponential terms for equation
(\ref{E-NewellWhiteheadSegel}) of the form $A^n\bar{A}^m
e^{iq(n-1-m)x}$. For the sake of simplicity we have consider the
dominant one in $A$, which correspond to the last term in Eq.
(\ref{E-NewellWhiteheadSegel}). Nevertheless, we expect that these
entire exponential small terms have qualitatively the same effect in
the dynamics. When the term proportional to spatial forcing is zero
($\eta \rightarrow 0$), the above model is the
Newell-Whitehead-Segel amplitude equation, which have been used
deeply to explain the appearance of stripe pattern \cite{Cross}.
Nevertheless, this model does not account for locking phenomenon
\cite{Pomeau}. The inclusion of spatial forcing terms in the
amplitude equations in 1D allows understanding the locking
phenomenon, the pinning range and localized structures
\cite{Clerc2005,Aranson2000,ClercFalcon2005}. In the extreme limit,
$\eta \rightarrow 0$, it is straightforward to show that model
(\ref{E-NewellWhiteheadSegel}) has a front solution connecting two
homogeneous states, $0$ and $\left( 1+\sqrt{1+4\epsilon }\right)
/2$, when $\epsilon <0$. Which accounts for the connection between
the stripe pattern and the uniform state. These two homogeneous
states are energetically equivalent at $\epsilon _{M}=-3/16$---the
{\it Maxwell point}---and it has the form
\begin{equation}
A=\sqrt{\frac{3/4}{1+e^{\pm \sqrt{3/4}(X-P)} }}e^{i\theta },
\label{E-FrontSolution}
\end{equation}
where $P$ stands for the position of flat interface, and $\theta $
is an arbitrary phase. This flat interface is transversally unstable
and it gives rise to a complex coarsening process, zigzag
instability. Numerically we have computed the growth rate of each
mode with wave number $k$ of the flat interface---spectrum.
Fig.~\ref{F-fig4}a depicts this instability and the spectrum of the
flat interface. Note that although the numerical simulations have
been done at the Maxwell point, the interface propagates from a
state that represents the stripe to a uniform one. When $\eta$ is
small the non-null uniform solution becomes a stripe state in the
y-direction (cf. Fig.~\ref{F-fig4}b). The zigzag dynamics exhibited
by the interface disappears and is replaced by a transversal
periodic structure, embroidery (cf. Fig.~\ref{F-fig4}b). Hence,
these transversal embroidery structures are consequence of the
interaction of the spatial forcing---generated by underlying
pattern---with the transversal instability of the
Newell-Whitehead-Segel model. Changing $\epsilon$ the embroidery
interface is motionless in a range of parameter of $\epsilon$, the
pinning range. Fig.~\ref{F-fig4}b shows the typical motionless
embroidery interface observed in model
(\ref{E-NewellWhiteheadSegel}) and the respective spectrum of the
flat interface. Increasing $\eta$ the embroidery interface becomes a
flat interface and the amplitude of the stripes increase.
Fig.~\ref{F-fig4}c shows the stable flat interface and its
respective spectrum. The flat and embroidery interfaces exhibit by
the amended amplitude equation are nonlinear unstable, that is, a
finite perturbation of these interfaces give rise a zigzag dynamics.
Therefore, the dynamics exhibits by the universal model
(\ref{E-NewellWhiteheadSegel}) is similar to those shown by
prototype model (\ref{E-SubcriticaSH}).

\section{Phenomenological 1D model}The model
(\ref{E-NewellWhiteheadSegel}) seems simpler than Eq.
(\ref{E-SubcriticaSH}), however, the analytical description of the
interfaces in these model is a thorny task. A standard method to
grasp the dynamics exhibited by the interface of the precedent
models is to derive a one-dimensional equation for it. This method
consists in using as an ansatz, the front solution
(\ref{E-FrontSolution}) plus a small correction, that
is,

\begin{equation}
A=\left\{ \sqrt{\frac{3/4}{1+e^{\pm\sqrt{3/4}(X-P(Y,\tau))}}}
+w_{0}(X-P,Y)\right\}  e^{i\theta},\label{E-FrontSolution}%
\end{equation}
where the continuous parameter $P$ is promoted to a field
($P\left( Y,\tau\right)  $), $w_{0}$ is a small complex correction
function, which is order of spatial variation of the position of
the interface ($w_{0} \sim\partial_{YY}P$)
\cite{Zigzag,Zigzag2,Calisto,Argentina}. However, we can not use
this weakly nonlinear method because the more unstable transversal
mode has a finite wave number (q). This type of method requires
that unstable modes have small wave number.  To
understand the mechanism of the different structures observed at
the interface, and based on symmetry arguments for the interface
\cite{Zigzag,Zigzag2,Calisto,Argentina} and on the
effect of spatial forcing term \cite{Clerc2005,Aranson2000}, we
propose the following phenomenological equation for the position
of the interface (\textit{convective and forced Cahn-Hilliard
equation})

\begin{equation}
\partial_{\tau}P=\varepsilon P_{YY}+P_{Y}^{2}P_{YY}-P_{YYYY}
+\alpha P_{Y}^{2}-\kappa\sin(QP)+\Delta \label{E-GCahn-Hilliard}
\end{equation}
this model has the equilibria $P_{n}=n\pi/Q$, $n=0$, $\pm1$, $\pm
2,\cdots$. The equilibria $P_{2m}$ as function of forcing $\kappa$
have similar spectrum to those shown in Fig.~\ref{F-fig4}. For large
$\kappa$, these equilibria are stable, i.e. this model has a family
of stable flat interfaces. Decreasing $\kappa$ these flat interfaces
become unstable and give rise to spatial periodical state,
embroideries. Finally, for small $\left\{ \alpha,\kappa\right\}  $
this model exhibit zigzag instability characterized by logarithmic
and power law for the coarsening. Hence, the one-dimensional model
(\ref{E-GCahn-Hilliard}) presents qualitative similar dynamics that
those shown by models (\ref{E-SubcriticaSH}) and
(\ref{E-NewellWhiteheadSegel}). In the coexistence region of uniform
states, one expects to find stable localized horm solution
\cite{ClercEscaffKenkre}, and then we expect to find localized horm
state at the interface. Fig.~\ref{F-fig5} illustrates the typical
stable horm solutions of one-dimensional model
(\ref{E-GCahn-Hilliard}) and the prototype model
(\ref{E-SubcriticaSH}). The study of the effect of the forcing on
zigzag dynamics of above model is in progress.

\section{Summary} Isotropic systems which have coexistence between stable
stripe pattern and uniform states can exhibit interfaces connecting
these states. These interfaces present a rich and unexpect
transversal dynamics like transversal spatial instability, nonlinear
zig-zag instability and localized states. Recently in Ref.
\cite{CEFRT}, it is shown that in anisotropic systems the flat
interface linking rolls pattern with uniform one is transversal
stable. Hence, the wealthy transversal dynamics is a consequence of
the spatial isotropy and the pinning effect.

Although the prototype model under study Eq.
(\ref{E-SubcriticaSH}) is variational---the dynamical evolution of
this model has the tendency to minimize its Lyapunov
functional---the interface dynamic exhibited by this model is
robust. Since, this interface dynamics is well described by the
amended amplitude equation, which is valid close to the spatial
instability. Hence, we expect that a system that exhibits a
coexistence between a stripe pattern and a homogeneous state
should be present a rich interface dynamics.

\acknowledgments The simulation software {\it DimX} developed at
INLN, France, has been used for all the numerical simulations. The
Authors acknowledge the support of ring program ACT 15 of
\textit{Programa Bicentenario de Ciencia y  Tegnolog\'ia} of Chilean
government. M. G. C. thanks the support of FONDAP grant 11980002. D.
E. thanks the financial support of Fondecyt 3070013 and FAI (Project
No. ICIV-003-08).  R. R. thanks the financial support of Fondecyt
3070039.


\begin{thebibliography}{0}

\bibitem{Prigogine}
    \Name{Nicolis G. \and Prigogine I.}
    \Book{Self-organization in NonEquilibrium systems}
    \Publ{J. Wiley and Sons, New York}
    \Year{1977}.

\bibitem{Cross}
    \Name{Cross M. C. \and Hohenberg P. C.}
    \Review{Rev. Mod. Phys.}
    \Vol{65}
    \Year{1993}
    \Page{851}.

\bibitem{Aranson}
    \Name{Aranson I. S. \and Tsimring L. S.}
    \Review{Rev. Mod. Phys.}
    \Vol{78}
    \Year{2006}
    \Page{641-692}.

\bibitem{NewellWhitehead}
    \Name{Newell A. C. \and Whitehead J. A.}
    \Review{J. Fluid. Mech.}
    \Vol{38}
    \Year{1969}
    \Page{279}.

\bibitem{Fisher}
    \Name{Fisher R. A.}
    \Review{Ann. Eugenics}
    \Vol{7}
    \Year{1937}
    \Page{335}.

\bibitem{Kolmogorov}
    \Name{Kolmogorov A., Petrovsky I. \and Piskunov}
    \Review{Bull. Univ. Moskou Ser. Int. Se.}
    \Vol{7}(6)
    \Year{1937}
    \Page{1}.

\bibitem{ClercEscaffKenkre}
    \Name{Clerc M. G., Escaff D. \and Kenkre V. M.}
    \Review{Phys. Rev. E}
    \Vol{72}
    \Year{2005}
    \Page{056217}.

\bibitem{Murray}
    \Name{Murray J. D.}
    \Book{Mathematical Biology},
    \Publ{Springer-Verlag, Berlin}
    \Year{1993}.

\bibitem{ClercResidori}
    \Name{Clerc M. G., Nagaya T., Petrossian A., Residori S. \and Riera C. S.}
    \Review{Eur. Phys. J. D}
    \Vol{28}
    \Year{2004}
    \Page{435}.

\bibitem{ResidoriPetrossian}
    \Name{Residori S., Petrossian A., Nagaya T., Riera C. S. \and Clerc M. G.}
    \Review{Physica D}
    \Vol{199}
    \Year{2004}
    \Page{149-165}.

\bibitem{Pomeau}
    \Name{Pomeau Y.}
    \Review{Physica D}
    \Vol{23}
    \Year{1986}
    \Page{3}.

\bibitem{Clerc2005}
    \Name{Clerc M. G., Falcon C. \and Tirapegui E.}
    \Review{Phys. Rev. Lett.}
    \Vol{94}
    \Year{2005}
    \Page{148302}.

\bibitem{Aranson2000}
    \Name{Aranson I. S., Malomed B. A., Pismen L. M. \and Tsimring L. S.}
    \Review{Phys. Rev. E}
    \Vol{62}
    \Year{2000}
    \Page{R5}.

\bibitem{MalomedNepo1991}
    \Name{Malomed B. A., Nepomnyashchy A. A. \and Tribelsky M. I.}
    \Review{Phys. Rev. A}
    \Vol{42}
    \Year{1990}
    \Page{7244}.

\bibitem{MeronElphick2006}
    \Name{Hagberg A., Yochelis A., Yizhaq H., Elphick C., Pismen L. \and Meron E.}
    \Review{Physica D}
    \Vol{217}
    \Year{2006}
    \Page{186}.

\bibitem{AhlersFront}
    \Name{Bodenschatz E., de Bruyn J. R., Ahlers G. \and Cannell D. S.}
    \Review{Phys. Rev. Lett.}
    \Vol{67}
    \Year{1991}
    \Page{3078}.

\bibitem{Taki}
    \Name{Durniak C., Taki M., Tlidi M., Ramazza P. L., Bortolozzo U., \and Kozyreff G.}
    \Review{Phys. Rev. E}
    \Vol{72}
    \Year{2005}
    \Page{026607}.


\bibitem{ClercFalcon2005}
    \Name{Clerc M. G. \and Falcon C.}
    \Review{Physica A}
    \Vol{356}
    \Year{2005}
    \Page{48}.

\bibitem{BrandSakaguchi}
    \Name{Sakaguchi H. \and Brand H. R.}
    \Review{Physica D}
    \Vol{97}
    \Year{1996}
    \Page{274}.


\bibitem{Budd}
    \Name{C.J. Budd \and R. Kuske.}
    \Review{Physica D}
    \Vol{208}
    \Year{2005}
    \Page{73}

\bibitem{Calisto}
    \Name{Calisto H., Clerc M., Rojas R. \and Tirapegui E.}
    \Review{Phys. Rev. Lett.}
    \Vol{85}
    \Year{2000}
    \Page{3805}.

\bibitem{Argentina}
    \Name{Argentina M., Clerc M. G., Rojas R. \and Tirapegui E.}
    \Review{Phys. Rev. E}
    \Vol{71}
    \Year{2005}
    \Page{046210}.

\bibitem{Zigzag}
    \Name{Chevallard C., Clerc M., Coullet P. \and Gilli J.-M.}
    \Review{Eur. Phys. J. E}
    \Vol{1}
    \Year{2000}
    \Page{179}.

\bibitem{Zigzag2}
    \Name{Chevallard C., Clerc M., Coullet P. \and Gilli J.-M.}
    \Review{Europhys. Lett.}
    \Vol{58}
    \Year{2002}
    \Page{686-692}.

\bibitem{CEFRT}
    \Name{Clerc M. G., Falcon C., Escaff D. \and Tirapegui E.}
    \Review{Eur. Phys. J. Special Topics}
    \Vol{143}
    \Year{2007}
    \Page{171}.

\end{thebibliography}
\end{document}